%Paper: hep-ph/9311361
%From: zack@yukawa.UCSD.EDU (Zachary S. Guralnik)
%Date: Wed, 24 Nov 1993 11:29:01 -0800

%%%%%%%%%%%%%%%%%%%%%%%%%%%%%%%%%%%%%%%%%%%%%%%%%%%%%%%%%%%%%%%%%%%
%                       INSTRUCTIONS
%
% This paper uses the harvmac macros. 3 files with postscript figures
% have been included as a uuencoded tar file with instructions for
% unpacking. If you have  epsf.tex, uncomment the following line
% and the postscript figures
%
%\input epsf
%
% will be included in the paper by the dvips program. If you do not
% have epsf.tex, you can print the figures out separately.
%
%%%%%%%%%%%%%%%%%%%%%%%%%%%%%%%%%%%%%%%%%%%%%%%%%%%%%%%%%%%%%%%%%%%%%%%%
%
%
\ifx\epsffile\undefined\message{(FIGURES WILL BE IGNORED)}
\def\insertfig#1{}% null macro
\else\message{(FIGURES WILL BE INCLUDED)}
\def\insertfig#1{{{
\midinsert\centerline{\epsfxsize=\hsize
\epsffile{#1}}\bigskip\bigskip\bigskip\bigskip\endinsert}}}
\fi
\input harvmac
%%%%%%%%%%%%%%%%%%%%%%%%%%%%%%%%%%%%%%%%%%%%%%%%%%%%%%%%%%%%%%%%%%%%%%
%
%  UCSD macros to overwrite some of the definitions in harvmac.tex
%  (include after harvmac.tex)
%  last modified 4/92
%
%%%%%%%%%%%%%%%%%%%%%%%%%%%%%%%%%%%%%%%%%%%%%%%%%%%%%%%%%%%%%%%%%%%%%%%
%
% modify the output routine for the little format
%
\ifx\answ\bigans
\else
\output={
  \almostshipout{\leftline{\vbox{\pagebody\makefootline}}}\advancepageno
}
\fi
%
%
% address
%

%
% grant numbers
%

%
% preprint number
%
\def\UCSD#1#2{\noindent#1\hfill #2%
\bigskip\supereject\global\hsize=\hsbody%
\footline={\hss\tenrm\folio\hss}}% restores pagenumbers
%
% abstract
%
\def\abstract#1{\centerline{\bf Abstract}\nobreak\medskip\nobreak\par #1}
%
%
% titlefont
%
%
\edef\tfontsize{ scaled\magstep3}
 \tfontsize  \tfontsize
 \tfontsize \font\titlei=cmmi10 \tfontsize
\font\titleis=cmmi7 \tfontsize \font\titleiss=cmmi5 \tfontsize
\font\titlesy=cmsy10 \tfontsize \font\titlesys=cmsy7 \tfontsize
\font\titlesyss=cmsy5 \tfontsize  \tfontsize
\skewchar\titlei='177 \skewchar\titleis='177 \skewchar\titleiss='177
\skewchar\titlesy='60 \skewchar\titlesys='60 \skewchar\titlesyss='60
%
%\def\titlefont{\def\rm{\fam0\titlerm}% switch to title font
%\textfont0=\titlerm \scriptfont0=\titlerms \scriptscriptfont0=\titlermss
%\textfont1=\titlei \scriptfont1=\titleis \scriptscriptfont1=\titleiss
%\textfont2=\titlesy \scriptfont2=\titlesys \scriptscriptfont2=\titlesyss
%\textfont\itfam=\titleit \def\it{\fam\itfam\titleit}\rm}
%
%
% math symbols
%
%---------------------------------------------------------------------
%
\def\inv{^{\raise.15ex\hbox{${\scriptscriptstyle -}$}\kern-.05em 1}}
  %prime
\def\lbar{{\lower.35ex\hbox{$\mathchar'26$}\mkern-10mu\lambda}} %lambda bar

%
%
% various slashed symbols
%
%
 % slashes a character
\def\dsl{\,\raise.15ex\hbox{/}\mkern-13.5mu D} %this one can be subscripted
\def\delsl{\raise.15ex\hbox{/}\kern-.57em\partial}
\def\Ksl{\hbox{/\kern-.6000em\rm K}}
\def\Asl{\hbox{/\kern-.6500em \rm A}}
\def\Dsl{\hbox{/\kern-.6000em\rm D}} %roman D
\def\Qsl{\hbox{/\kern-.6000em\rm Q}}
\def\gradsl{\hbox{/\kern-.6500em$\nabla$}}
%
% space and backspace in l mode
%
\def\lspace{\ifx\answ\bigans{}\else\qquad\fi}
\def\lbspace{\ifx\answ\bigans{}\else\hskip-.2in\fi} % $$\lbspace...$$
%
%     boxes an equation
%
\def\boxeqn#1{\vcenter{\vbox{\hrule\hbox{\vrule\kern3pt\vbox{\kern3pt
        \hbox{${\displaystyle #1}$}\kern3pt}\kern3pt\vrule}\hrule}}}
%
%     draw a little box (end of proof symbol)
%     e.g. \mbox{.1}{.1}
%
\def\mbox#1#2{\vcenter{\hrule \hbox{\vrule height#2in
\kern#1in \vrule} \hrule}}
%
%
%
%     curly letters
%
   %curly letters

%
%
%
%     derivatives
%
%
\def\del{\partial}

\def\bar#1{\overline{#1}}

\def\bra#1{\left\langle #1\right|}
\def\ket#1{\left| #1\right\rangle}

\def\darr#1{\raise1.5ex\hbox{$\leftrightarrow$}\mkern-16.5mu #1}

%
 %pound sterling
%
 %puts a small half in a displayed eqn
\def\frac#1#2{{\textstyle{#1\over #2}}} %puts a small fraction
%in a displayed eqn
%
%
%     various math operators
%
%

%
%
%
%

%
%       relations
%
\def\ltap{\ \raise.3ex\hbox{$<$\kern-.75em\lower1ex\hbox{$\sim$}}\ }
\def\gtap{\ \raise.3ex\hbox{$>$\kern-.75em\lower1ex\hbox{$\sim$}}\ }
\def\gl{\ \raise.5ex\hbox{$>$}\kern-.8em\lower.5ex\hbox{$<$}\ }
\def\roughly#1{\raise.3ex\hbox{$#1$\kern-.75em\lower1ex\hbox{$\sim$}}}
%
%
%       This defines et al., i.e., e.g., cf., etc.

%

%

\relax

\noblackbox
\def\JJQ{\bra{0}T^{\ast}\left( J^{\mu}_5 (x)\ J^{\alpha}(y)\right) \ket{0}}
\def\JJ{\bra{0}T^{\ast} \left( J^{\mu}(x)\ J_5^{\alpha}(y)\right) \ket{0}}
\def\del{\partial}
\def\kl{\hat K \hat K ^{\dagger}}
\def\kr{\hat K ^{\dagger} \hat K}
\centerline{{\titlefont{Anomalies without Massless Particles}}}
\bigskip
\centerline{Zachary Guralnik}
\smallskip
\centerline{{\sl Department of Physics, University of California at San
Diego, La Jolla, CA 92093}}
\bigskip
\vfill
\abstract{Baryon and lepton number in the standard model are violated
by anomalies, even though the fermions are massive.  This problem
is studied in the context of a two dimensional model.  In a
uniform background field,  fermion production arise from non-adiabatic
behavior that compensates for the absence of massless modes.  On
the other hand, for localized instanton-like configurations, there
is an adiabatic limit.  In this case, the anomaly is produced by
bound states which travel across the mass gap. The sphaleron
corresponds to a bound state at the halfway point.}

\vfill
%\draftmode
\UCSD{\vbox{\hbox{UCSD/PTH 93-33}\hbox{hep-ph/9311361}}}{November 1993}

\newsec{Introduction}
The divergence of lepton and baryon currents in the Standard
Model is independent of the fermion masses.
For a single family, the baryon and lepton number anomaly
is
\eqn\bnm{
\del_{\mu}J^{\mu}_B=\del_{\mu}J^{\mu}_L=
{g^2\over 32\pi^2}
\epsilon_{\mu\nu\alpha\beta} {\rm Tr} (W^{\mu\nu}W^{\alpha\beta})
-{g^{\prime 2}\over 64\pi^2}\epsilon_{\mu\nu\alpha\beta}
B^{\mu\nu}B^{\alpha\beta}\ ,}
where $W^{\mu\nu}$ is the $SU(2)$ field strength and $B^{\mu\nu}$
is the $U(1)$ field strength.
This differs greatly from the axial current equations
of Q.E.D. because in Q.E.D.
the production of axial charge depends critically on whether or not
the electron is
massive.  I will begin by reviewing the reasons for this
sensitivity.  Then I will show why these reasons are not applicable to a
spontaneously broken theory with a vector current anomaly, such as
the standard model.  The results give some insight into the production
of baryon number in the standard model by sphalerons,  which has been
of much recent interest.

The divergence of the axial current in Q.E.D.~\ref\jack{S.L. Adler,
Phys. Rev. 177 (1969) 2426\semi
J.S. Bell and R.Jackiw, Nuovo Cimento 60A (1969) 47.}
is
\eqn\axnum{\del_{\mu}\bar\psi\gamma^{\mu}\gamma^5
\psi={e^2\over 16\pi^2}\epsilon_{\mu\nu\alpha\beta}
F^{\mu\nu} F^{\alpha\beta}
+2im\bar\psi\gamma^5\psi.}
In a background gauge field the matrix element of the last term is
\eqn\canc{_A\bra{0\, out}2im\bar\psi\gamma^5\psi
\ket{0\, in}=-{e^2\over 16\pi^2}
\epsilon_{\mu\nu\alpha\beta}F^{\mu\nu}F^{\alpha\beta}
+\dots}
The remaining terms are higher dimension functions of the gauge
fields and vanish in an adiabatic aproximation.
If the electron is massive then there is no axial charge
violation in an adiabatic approximation because the first and last
terms in equation~\axnum\  cancel.
This cancellation is obvious from the start if one
calculates the anomaly using a Pauli Villars
regulator field.  Then the regulated axial current satisfies
\eqn\vill{\del_{\mu}J^{5\mu}_r=2i\Lambda\bar\chi\gamma^5\chi+
2im\bar\psi\gamma^5\psi\ ,}
where $\chi$ is the regulator field and $\Lambda$ is its mass.
$\chi$
is bosonic, so $\chi$ loops have the opposite sign from $\psi$ loops.
Therefore there can be no mass independent terms in the matrix element
of $\del_{\mu}J^{5 \mu}_r$ in a background gauge field.

This cancellation also has a simple
spectral interpretation.  An
explanation of the Q.E.D. axial anomaly based upon the spectrum of
a massless electron in a background magnetic field has been given
by Nielson and Ninomiya~\ref\who{H.B. Nielsen and M. Ninomiya,
Int. J. Mod. Phys A6 (1991) 2913.}.
Their arguments are briefly summarized below.
Consider
a uniform background magnetic field in the
z direction.  In the massless case, positive and negative chirality
fermions  decouple, so
there are two sets of Landau levels.
The positive and negative chirality Landau levels contain zero-modes
with $E=-p_z$ and $E=+p_z$ respectively.
Suppose one turns on a positive
uniform electric field ${\cal E}$ in the $z$ direction.
In an adiabatic approximation, solutions flow along spectral lines
according to the Lorentz force law ${dp\over dt}=e{\cal E}$.  Thus
right chiral zero-modes slide out of the Dirac sea while
left chiral zero-modes slide deeper into the Dirac sea (\fig\figone{
Illustration of the positive and negative chirality
spectral flows of massless $3+1$ Q.E.D. which
produce the axial anomaly.  The solid lines indicate occupied states,
while the dashed lines indicate empty states.}).  This
motion produces a net axial charge but no electric charge.
By a careful counting of states one reproduces the global form
of the anomaly
\eqn\gob{{dQ^5\over dt}={\rm V}{e^2\over 2\pi^2}{\cal E}_z{\cal B}_z\ ,}
where V is the volume of space.  Now consider the same background
fields but suppose the electron is massive.
In this case, there are no zero-modes among the Landau levels.
In the absence of zero-modes  adiabatic evolution just maps
the Dirac sea into itself, so axial charge can not be
adiabatically generated.

The discussion above is not applicable to the standard model because
standard model fermions can be given masses without changing
the baryon or lepton number violation in  fixed gauge field
background.  Dirac mass terms do not carry
vector charge, so they do not effect the divergence of a
vector current.  Yet in an adiabatic limit it
seems that presence or absence of mass terms $\it must$ effect the
divergence of a current.
In the following, this paradox will be resolved by
solving the
equations of motion for certain background fields which, according
to the anomaly equation, should generate charge.  I will demonstrate
that spatially uniform backgrounds which generate
vector charge
have no adiabatic limit.  Such backgrounds produce the anomaly
by causing hopping between energy levels. On the other hand, localized
instanton-like backgrounds
do possess an adiabatic limit.  Backgrounds of this type will be shown
to produce the anomaly via  fermionic bound states whose energies
traverse the gap between $E=-m$ to $E=m$.  This give a better
understanding of the mechanism of baryon number production in the
standard model by sphalerons.  The sphaleron configuration corresponds
to the half-way point with a zero energy bound state.

Because of the chiral couplings, the standard model Landau levels
are quite complicated.  To avoid calculating Landau levels in
$3+1$ dimensions,  I will
instead consider a spontaneously broken $U(1)$
axial gauge theory in $1+1$ dimensions.
While the details of the computation are different, many of the results
obtained in $1+1$ dimensions are expected to hold in $3+1$ dimensions.
The lagrangian of this theory is
\eqn\full{
{\cal L}=-{1\over 4g^2}F^{\mu\nu}
F_{\mu\nu} + \bar\psi (i\delsl + \Asl \gamma^5
-\lambda\phi^*P_L - \lambda\phi P_R)\psi + {1\over 2}
D^{\mu}\phi^{\ast}D_{\mu}\phi - U(\phi^*\phi)\ .}
This simplified model possesses the two traits whose
consistency I wish to demonstrate;
a massive spectrum and a mass independent vector
current divergence,
\eqn\cog{\del_{\mu}\bar\psi\gamma^{\mu}\psi={g\over 2\pi}\epsilon
_{\mu\nu}F^{\mu\nu}\ .}
For the moment I will not
consider the full dynamical theory,  but only that given by
\eqn\lag{{\cal L} =
\bar\psi (i\delsl+ g\Asl(x)
\gamma^5- \lambda \rho(x) e^{i\theta(x)\gamma^5})\psi\ ,}
where $\rho(x) =v$ asymptotically.
It should be possible to demonstrate the anomaly
by considering the momentum space equations of motion,
as was done for massless Q.E.D. by Nielsen and Ninomiya
using the Lorentz force law.
A few remarks are in order about how to do this.
Let the Dirac field in a background be expanded as follows:
\eqn\expan{\psi (x,t)= \int {dp\over 2\pi} c_{p,i}(t)
u_{p,i}e^{ipx}\ ,}
where $u_{p,i}$ are free massive spinors
normalized to 1,  and the index $i$ distinguishes
between positive and negative
frequency solutions when the backgrounds vanish.
All the background dependance is contained in the time evolution of
$c_{p,i}(t)$
When the backgrounds vanish, \eqn\vanish{c_{p,i}(t)=
\exp(i\omega_{p,i}t) c_{p,i}(0)\ ,}
where \eqn\megr{\omega_{p,\pm}=\pm\sqrt{p^2+m^2}\ .}
Given a knowledge of which states
are occupied at an initial time,  one can determine which states
are occupied at a final time by looking at the evolution of the
coeficients $c_{p,i}$.
At this point however, the use of this expansion
to determine the vector charge or the particle number
is very ambiguous.  One can make
transformations of $\psi$, corresponding to certain transformations
of the background fields, which change the $c_{p,i}$.  For example
transformations exist which map something that
looks like the Dirac sea into something that looks like an
excited state with non zero vector charge.
An  invariant definition of charge is needed.
Such a definition must depend on the background fields as well as
the Fourier coefficients.
In order to make the computation of the charge simple,  I
will only consider processes in which local gauge invariant
functions of the
background fields vanish at asymptotic times.
This means that the initial and final
$\theta$ and $A^{\mu}$
are gauge equivalent to $\theta =0$ and $A^{\mu} =0$.  In this case
the proper definition of charge  at asymptotic times is simple.
In Dirac sea language, one subtracts the number of vacant
negative frequency states from the number of occupied positive
frequency states.
The occupation number of a positive
or negative frequency state of momentum $p$ is proportional to
$|c_{p,\pm}|^2$ in the gauge in which
the backgrounds vanish.
Equivalently, in second quantized language one can
adopt a normal ordered definition of charge at asymptotic times.
The change in the charge can then be written in terms of
Bogolubov coefficients relating the operators $\hat c_{p,i}$
in the asymptotic past to those in the asymptotic future, where
these operators are defined in the gauge in which
the backgrounds vanish.
Note that at intermediate times  the gauge invariant backgrounds
do not vanish so a well defined Bogolubov transformation
between asymptotic past and intermediate times does
not exist.  Normal ordering is no longer
sensible at intermediate times because solutions can not be
classified as positive
or negative frequency.
However,  I will never explicitly calculate
the charge at intermediate times \footnote*{At intermediate times the
charge is defined by axial gauge invariance and charge conjugation
symetry.
For example one can use an axially gauge invariant point split
charge which is odd under charge conjugation.  When the gauge fields
vanish this is equivalent to the usual normal ordered definition of charge.}.

\newsec{Uniform Backgrounds}

In the spirit of the anomaly calculations done by Nielsen and
Ninomiya,
I will first consider a process in which a spatially uniform axial electric
field is turned on and then off.  I will also choose a uniform
(spatially parallel transported) Higgs field
background. The particular background to be considered is
\eqn\backg{A^0=- {\cal E}(t)x,\quad A^1 =0,\quad\theta = 0\ ,}
where ${\cal E}(t)={\cal E}$ for $0<t<T$ and $0$ at all
other times.  In this gauge, with the initial and final backgrounds
vanishing,  the coefficients $c_{p,i}$ have an immediate interpretation
in terms of particle and charge production.
Due to the
axial electric field, vector
charge generation is expected, and should be evident in
the time evolution of these coeficients.  The equations of
motion for $c_{p,i}(t)$ are complicated at low p, but simplify
greatly at large $|p|$.  The simplification occurs because, as one
would expect,  the fermion mass can be neglected at large
$|p|$. A straightforward calculation gives
the equation describing behavior deep in
the Dirac sea:
\eqn\larp{\eqalign{{d\over dt} c_{p,-}(t) =& {gE\over\pi}
u^{\dagger}_{p,-}\gamma^5 u_{p,-}{d\over dp}c_{p,-}(t)
-i\omega_{p,-}c_{p,-}(t)\ ,\cr
=&{p\over |p|}{gE\over\pi}{d\over dp}c_{p,-}(t)
-i\omega_{p,-}c_{p,-}(t)\ .\cr}}
This equation is not complete,  but the neglected terms are all
supressed by factors of ${m\over |p|}$.
The solution is
\eqn\triv{c_{p,-}(t)=c_{\tilde p,-}(0)\exp(-i\omega_{p,-}t)\ ,}
where \eqn\cud{\tilde p=p-{p\over |p|} {gE\over\pi}t\ ,}
which is easily recognized as an axial version of the Lorentz force
law.
Therefore states along the
negative frequency spectral lines at large $|p|$ flow inward towards
small $|p|$.  Because of  unitarity and Fermi statistics, solutions
can not pile up at small $|p|$.
Therefore there
must be level
hopping at small $|p|$.
Positive frequency states must appear at a rate matching the
inward flow of negative frequency states across some large
$|p|$ cutoff (\fig\figtwo{Illustration of the spectral
flows which produce the vector current
anomaly in a spontaneously broken two dimensional axial gauge theory.}).
I thus arrive at the result that the backgrounds of~\backg\
have no adiabatic limit. Therefore
the absence
of zeromodes has no effect on charge production.
Putting
the system on a line of length L with periodic boundary boundary
conditions on the Fermi field,
one finds that the number of states crossing the cutoff
per unit time is ${L\over\pi}gE$.  This yields the
expected anomaly ${1\over L}{dQ\over dt}= {gE\over \pi}$.

There is actually no reason to expect adiabatic behavior with
uniform backgrounds.
The backgrounds of~\backg\  have singular time dependence when the electric
field is turned on or off.  One can make the time dependence of
these backgrounds nonsingular
either by smoothly switching the electric field
on and off,  or by going to $A^0=0$ gauge.  If one does the former,
one can try to make the backgrounds vary slowly in time by having
the electric field ${\cal E}(t)$ vary slowly in time.
However, no matter how slowly
the electric field varies,  $A^0$ will vary rapidly at large distances
since $A^0=-{\cal E}(t)x$.
In $A^0=0$ gauge,  the backgrounds of \backg\ become
\eqn\crap{\eqalign{A^1=&0,\quad\theta =0\cr
                  A^1=&{\cal E}t,\quad\theta = -2{\cal E}xt\cr
         A^1=&{\cal E}T,\quad\theta = -2{\cal E}xT\cr}\qquad {\rm for}\qquad
        \eqalign{&t<0\ ,\cr
                 0<&t<T\ ,\cr
                 T<&t\ .\cr}}
One can try to make these backgrounds vary slowly in time by
making ${\cal E}$ small.
Yet,  no matter how small ${\cal E}$ is,  the Higgs phase $\theta$ winds
wildly with time at large distances.  Therefore the non adiabatic nature of
uniform charge producing backgrounds is an infinite
volume effect.

It is actually easy to see the low momentum level hopping
explicitly without
invoking Fermi statistics.  In $1+1$ dimensions
$\gamma^{\mu}\gamma^5 =\epsilon^{\mu\nu}\gamma_{\nu}$.
One can use this fortuitous fact to solve the equations of motion
at all momenta.  For $0<t<T$
the background fields of ~\backg\  are equivalent to a
a background vector gauge field~\footnote*{This method of solving
the Dirac equation brings up a troubling question.  If an axial
gauge field background can generate vector charge, then
apparently a vector gauge field background can also generate
vector charge.  I discuss why this last statement is
not true in appendix A.}
with $V^0=0$ and $V^1=-{\cal E}x$.
The vector field strength vansishes, so the time
evolution of $\psi$ at intermediate times is trivial.
$\psi^{\prime}\equiv e^{{i\over 2}{\cal E}x^2}\psi$ evolves as a free
field:
\eqn\free{c^{\prime}_{p,i}(t)=e^{-i\omega_{p,i}t}
c^{\prime}_{p,i}(0)\ ,}
One only has to transform back from $c^{\prime}$ to $c$ to
get $c_{p,i}(t)$ as a function of the initial coefficients $c_{r,l}(0)$.
The result is that
\eqn\mess{c_{p,i}(t)=\sum_l \int {dr\over 2\pi}T_{p,i;r,l}
c_{r,l}(0)\ ,}
where
\eqn\more{T_{p,i;r,l}={2\pi\over E}u^{\dagger}_{p,i}
\left[\sum_j \int dq \exp(i{p-r\over E}q-i\omega_{q,j}t)
u_{q,j}u^{\dagger}_{q,j} \right]u_{r,l}
\exp(-i{p^2-r^2\over 2E})\ .}
The quantity within the brackets can be written
\eqn\amaz{{i\over2}
\gamma^0 (i\delsl -m)\bra{0}\left[\phi (z),\phi (0)
\right]\ket{0}\ ,}
where $\phi$ is a massive free scalar field in $1+1$ dimensions,
and $(z^0,z^1) \equiv (t,{p-r\over E})$.
The ``light cone'' singularity in~\amaz\ gives the leading term
of~\more\ :
\eqn\lead{
T_{p,i;r,l}= {\pi\over E}\bar u_{p,i}\left( i\gamma^+ \delta (z^+)
+i\gamma^- \delta (z^-)\right)u_{r,l}\exp(-i{p^2-r^2\over 2E})+\dots}
Let us rewrite this in a form which is easier to interpret:
\eqn\rew{\eqalign{T_{p,i;r,l}= {2\pi \over E}&
\left[ u^{\dagger}_{p,i} {1-\gamma^5\over 2}
u_{r,l}\delta (t+{p-r\over E})+
u^{\dagger}_{p,i} {1+\gamma^5\over 2}
u_{r,l}\delta (t-{p-r\over E})\right] \cr
\times& \exp(-i{p^2-r^2\over 2E}) \cr}\ ,}
where I have used the fact that
$\gamma^0\gamma^{\pm}=1 \pm
\gamma^5$ in  $1+1$ dimensions.
The axial Lorentz force law is clearly visible in the delta functions
and the associated left or right chiral projectors.
The low momentum level hopping is also manifest.
The hopping of negative frequency to positive frequency
states is described by $T_{p,+:r,-}$.
At large $p$ and $r$ of the same sign,  the spinors $u_{p,+}$
and $u_{r,-}$ have opposite chirality so that $u^{\dagger}_{p,+}
{1\pm\gamma^5\over 2}u_{r,-}$ vanishes.  Thus in the limit of large
momenta at fixed time, $T_{p,+;r,-}$ vanishes.
However at small $|p|$ the spinors have mixed chirality so
that~\rew\ does not vanish when $p-r=\pm {\cal E}t$, and
the predicted level hopping occurs.
It is interesting to note that
factor ~\rew\ is almost
the transformation function associated with an axial transformation of
$\psi$:
\eqn\axt{\psi (x)\rightarrow \psi^{\prime}(x)=\exp(-iEtx
\gamma^5)\psi (x)\ ,}
is equivalent to
\eqn\fourier{
c_{p,i}\rightarrow
c^{\prime}_{p,i}=\sum_l \int {dr\over 2\pi}T^{\prime}_{p,i;r,l}c_{r,l}
\ ,}
where \eqn\veq{T^{\prime}_{p,i;r,l}=T_{p,i;r,l}\exp(-i{p^2-r^2\over 2E})\ .}
This is not to be confused with an axial $\it gauge$ transformation
because the initial and final background fields are the same;
$A^{\mu}=0$ and $\theta=0$.  An axial gauge
transformation does nothing,
but an axial transformation which leaves the Higgs and gauge potentials
unchanged can produce particles and vector charge.  This should be
no surprise given the bosonization rules~\ref\bos{S. Coleman,
Phys. Rev. D11 (1975) 2088\semi
S. Mandlestam, Phys. Rev. D11 (1975) 3026.}
for an axial gauge theory
in $1+1$ dimensions.  The vector charge density in bosonized form
is \eqn\boz{J^0={1\over\sqrt{\pi}}(\del_1\chi-{1\over\sqrt{\pi}}A^1)\ ,}
where $\chi$ is the bosonic counterpart to $\psi$.  An axial
transformation \eqn\cor{\psi\rightarrow e^{if(x)\gamma^5}\psi\ ,}
corresponds to \eqn\crsp{\chi\rightarrow \chi+{1\over\sqrt{\pi}}f(x)\ .}
Therefore an axial transformation of the type~\axt\ above produces a net
vector charge.

\newsec{Localized Backgrounds}

The uniform backgrounds of~\backg\ are interesting but perverse because
the gauge invariant
objects built from the Higgs and gauge fields do not fall off
at large spatial distances.
Furthermore
these configurations can exist only in an infinite volume because
they are inconsistent with periodic boundary conditions.
Therefore let us instead consider localized, charge producing backgrounds.
By localized,  I mean that the energy density carried by
the backgrounds is at its minimum outside a spacetime disc of finite radius.
At fixed $\Delta Q$
one can always make such
backgrounds vary arbitrarily slowly in time, so that there is no
argument against the existence of an adiabatic limit.
We are again confronted with the puzzle
of how vector charge can be
produced by a weak electric field in a theory with a gap.

The clue to the puzzle is that one can not go to unitary
$(\theta=0)$ gauge from localized backgrounds which produce charge.
For such backgrounds $D_{\mu}\phi=0$ asymptotically.  Therefore
\eqn\another{\oint dx^{\mu}\del_{\mu}\theta=-2g\oint dx^{\mu}A_{\mu}=
-2\pi\Delta Q\ .}  If $\Delta Q$ is not zero, then $\phi^{\ast}\phi$
must vanish somewhere due to the non vanishing Higgs winding number.
In the presence of such a defect there
may be a bound state as
well as the continuum of ``scattering''
solutions with $E=\pm\sqrt{p^2+m^2}$.
In an adiabatic limit the only way charge can appear is if a
bound state traverses the mass gap.
As the defect is created and
destroyed in a process with $\Delta Q=1$, the bound
state energy should change continuously from $-m$ to $m$.
I will show that this is indeed the case.
The sphaleron
corresponds to a bound
state at the half-way
point and has charge one half~\ref\rorsh{R. Jackiw and C. Rebbi,
Phys. Rev. D13 (1976) 3398\semi
J. Goldstone and F. Wilczek, Phys. Rev. Lett. 47 (1981) 986\semi
F. Klinkhamer and N. Manton, Phys. Rev. D30 (1984) 2212.}.

An example of a localized configuration giving $\Delta Q =1$
is
\eqn\Higgs{\eqalign{\phi=& v\exp\left( i\alpha (t){x\over |x|} \right)\ ,\cr
                    A^0=&-{1\over 2g}{x\over |x|}{d\alpha\over dt}\ ,\cr
                    A^1=&0\ ,\cr}}
where the phase $\alpha (t)$ rotates by a total angle of  $-\pi$ from
$\alpha (-\infty)=0$ to $\alpha (\infty)=-\pi$.
In an adiabatic limit $\alpha(t)$ varies slowly and the
gauge fields can be neglected.
The defect at $x=0$ is spatially pointlike for convenience;
For a fixed $\alpha$,  finding  the spectrum is a trivial
matching problem.  (A less singular version
of this background is drawn in \fig\figthree{A winding higgs field
background for $\Delta Q=1$.
The time axis is vertical and the space axis
is horizontal.})
One finds a set of scattering solutions with
$E=\pm\sqrt{p^2+m^2}$,  but there is also a bound state solution
with $E^2<m^2$.  Continuity of the solution across $x=0$ requires
\eqn\eigs{e^{-2i\alpha}={E+i\sqrt{m^2-E^2}\over
E-i\sqrt{m^2-E^2}}\ .} This yields a bound state with energy
$E=-m\cos\alpha$.
As $\alpha$ varies adiabatically from $0$ to $-\pi$,
a single bound charge is carried across the gap.
Note that this alone does not guarantee the net production of charge.
A bound state could travel across the gap and leave a negative energy
hole.  The axial Lorentz force law causes negative frequency states
to slide inwards
towards zero momentum, which prevents the appearance of a hole.
In an adiabatic approximation, the gauge fields are negligible pertubations
on the spectrum,  but
drive the spectral flows needed to produce the anomaly.

For more general localized backgrounds,
an index theorem enables one to count the number of
time dependent energy eigenvalues
which travel across the gap.
Consider spinor functions $f(x,\tau)$ anihilated by the operator
\eqn\dd{\hat D\equiv {\del\over\del\tau}+\hat H (\tau)\ ,}
where by varying the parameter $\tau$  from
$-\infty$ to $\infty$ one goes slowly through
the same cycle of Dirac hamiltonians $\hat H$ that occur
in real time.  I will write the energy eigenvalues as
$E_n(\tau)$ and the energy eigenfunctions as $\chi_n (x,\tau)$.
Since $\hat H (\tau)$ is a slowly varying function of $\tau$,
the solutions of equation~\dd\  can be
written as
\eqn\slns{f(x,\tau)=a_n(\tau)\chi_n(x,\tau)\ ,}
where there is no sum on $n$ and\eqn\dcy{a_n(\tau)=
a_n(0)e^{-\int_0^{\tau}d\tau^{\prime}E_n(\tau^{\prime})}\ .}
This solution is only normalizable
if $E_n(\tau)$ has a negative value at
$\tau=-\infty$ and a  positive value at $\tau=+\infty$.
Now consider the adjoint operator \eqn\adjnt{\hat D^{\dagger}
\equiv -{\del\over\del\tau}+H(\tau)\ .}  A function $a_n(\tau)
\chi_n(x,\tau)$ annihilated by $\hat D^{\dagger}$ is only normalizable if
$E_n(\tau)$  has  a positive value at $\tau=-\infty$ and a
negative value at $\tau=+\infty$.  Hence the total charge
generated by bound states crossing the gap is equal to the
difference in the number of normalizable modes annihilated by
$\hat D$ and the
number of normalizable modes annihilated by
$\hat D^\dagger$\footnote*
{Witten has applied similiar
methods to
a different problem~\ref\witten{E. Witten,  Phys. Lett. 117B (1982) 324.}.}.
This quantity is
known as the index of $\hat D$.
The operator whose
index I wish to calculate is
\eqn\ours{\hat D ={\del\over\del\tau}+\gamma^0\left( i\gamma^1
(\del_1+igA_1\gamma^5)-\lambda\phi(x,\tau)
{1+\gamma^5\over 2}-\lambda\phi^{\ast}
(x,\tau){1-\gamma^5\over 2}\right)\ ,}
where asymptotically \eqn\asymp{\phi=ve^{i\theta}\ .}
$A^0$ is absent from $\hat D$ because it is
negligible in an adiabatic approximation.
One can take the adiabatic limit
of a process with fixed $\Delta Q$ by making
the following gauge invariant rescaling of the fields:
\eqn\rescal{\eqalign{\phi^{\prime}(x,t)
=&\phi(x,{t\over\lambda})\ ,\cr
A^{\prime 0}(x,t)=& {1\over\lambda}A^0(x,{t\over\lambda})\ ,\cr
A^{\prime 1}(x,t)=& A^1(x,{t\over\lambda})\ .\cr}}
In the large $\lambda$ limit $A^0$ vanishes.
$A^1$
is a nonvanishing adiabatic parameter, but one can gauge it to zero.
Doing so effects only the eigenfunctions of $\hat H (\tau)$ but not
the eigenvalues.
A straight-forward method to calulate the index of Dirac operators
on $R_n$ has been constructed by
Weinberg\ref\wein{E.J. Weinberg, Phys. Rev. D24
(1981) 2669.}.
Using these methods, the index of $\hat D$ with $A^1=0$
is found to be \footnote*{Weinberg applied his methods to count the
number of zero energy modes of a vortex-fermion system in
2 spatial dimensions.
This system was previously considered by Jackiw
and
Rossi~\ref\rossi{R. Jackiw and P. Rossi, Nucl. Phys. B190 (1981) 681.}
who suggested
the existence of an index theorem
equating the number of fermion zero energy modes to the
vortex number.  The index theorem for their model is very similiar
to the one considered in this paper.}
\eqn\idex{{1\over 2\pi}\oint dx^{\mu}\del_{\mu}\theta\ ,}
which is gauge invariant.
This is just as one expects given equation~\another .

The relation of this index theorem to charge production
can also be understood in terms of the euclidean path integral
using methods due to Fujikawa~\ref\fujj{K. Fujikawa, Phys. Rev. Lett. 42
(1979) 1195.}\ and
't Hooft~\ref\thooft{G. 't Hooft, Phys. Rev. Lett.
37 (1976) 8.}.
The fermionic portion of the partition function is
\eqn\eucaction{\int {\cal D}\psi {\cal D}\bar\psi
\exp\left( -\int d^2x\bar\psi {\hat K} \psi \right)\ ,}
where
\eqn\kdef{K=\gamma^0(\del_0-{\hat H}).}
Let $\psi$ and $\bar\psi$ be expanded as
\eqn\epand{\eqalign{\psi(x) =& \sum_n a_n f_n(x)\ ,\cr
                    \bar\psi(x) =& \sum_l {\bar b}_l g_l^{\dagger}(x)\ ,\cr}}
where
\eqn\where{\eqalign{\hat K ^{\dagger} \hat K f_n(x) =&
              \lambda_n f_n(x)\ , \cr
               \hat K \hat K ^{\dagger}g_l(x) =& \alpha_l
               g_l(x)\ , \cr}}
and $f_n(x)$ and $g_l(x)$ are normalized to one.
There is a one to one mapping between eigenfunctions of
$\hat K ^{\dagger} \hat K$ and $\hat K \hat K ^{\dagger}$ provided
that the eigenvalue is not zero.  $\hat K$ maps eigenfunctions of
$\kr$ into eigenfunctions of $\kl$ with the same non zero eigenvalue,
while
$\hat K ^{\dagger}$ does the inverse mapping.  However if $\kr f(x) =0$
or $\kl g(x) =0$,  then there is no mapping because $\kl f(x)=0$
implies that $\hat K f(x) =0$, and $\kr g(x)=0$ implies
that $\hat K^{\dagger} g(x)=0$.
The difference between the
number of zeromodes of $\kr$ and $\kl$ is given by the index of $\hat K$.
A zeromode of either $\kr$ or $\kl$ contributes
nothing to the euclidean action.  Therefore the integral over
the grassman coefficient of a zeromode will vanish unless the
coefficient appears in the expansion of an operator in a Green's function.
It is easy to see from this that the contributions of a given Higgs and
gauge field background to a Green's function vanishes except when the
number of $\psi$'s in the Green's function differs from the number
$\bar\psi$ 's by the index of $\hat K$. For example, if $\kr$ has
one zeromode $f_0(x)$ and $\kl$ has no zeromode, then
\eqn\thoft{\int {\cal D}\psi {\cal D}\bar\psi\psi(x)
\exp\left(-\int d^2y\bar\psi\hat K \psi \right)=\sqrt{det \kl}f_0(x)\ .}
In general the net vector charge produced is
given by the index of $\hat K$, which in an adiabatic limit is the same
as
the index of
$\hat D$ because
the two operators differ only by a factor of $\gamma^0$.
The connection between the spectral and path integral approaches to the anomaly
is now clear~\footnote*{This
connection is not novel.  The relation between modes
annihilated by the Euclidean Dirac operator and
spectral flows which take states in and out of the Dirac sea was discussed
by Nielsen and Ninomiya in the context
of massless fermions~\who . }.

An interesting feature of the index theorem for a
spontaneously broken axial theory is that it permits Higgs and gauge
field
backgrounds to create single
fermions and not just pairs.
The Euclidean equations of motion possess a symetry
$\psi\rightarrow\gamma^0\psi^{\ast}$.
In the absence of the Higgs coupling to fermions,
$\hat K$ anticommutes with $\gamma^5$, so zeromodes can be chosen to be
chiral.
Therefore in the massless axial theory
zeromodes occur in pairs of opposite chirality which are related by
the above symetry.  This pairing is a reflection of $Q_5$
conservation.
However in the spontaneously
broken axial theory,  $Q_5$ has a Higgs component as well as a fermionic
component, and only the sum is conserved.  It is no longer true that
$\lbrace\hat K , \gamma^5\rbrace = 0$. Therefore zeromodes can no longer
be chosen to be chiral.
In fact,  in an adiabatic
approximation one can prove that
the mapping $\psi\rightarrow\gamma^0\psi^{\ast}$
does not yield independent solutions.  This is done in
appendix B.  The production of single fermions by a background is not
a violation of gauge or Lorentz invariance.  For example
a single fermion can not
get a vacuum expectation value
because the path integral over gauge and Higgs
fields in the one instanton sector vanishes, even if the fermionic
integral does not.

\newsec{Dynamics}

So far it has only been demonstrated
how charge violation proceeds independently of
the fermion masses in the case of background Higgs and gauge
fields.  I will now show how this works in the dynamical case.
This will be done by demonstrating the consistency of the
Ward identities
with a massive spectrum.  Similiar results should hold for
three current correlation functions in $3+1$ dimensions.

The current equations are
\eqn\current{\del_{\mu}J^{\mu}_5
=\del^{\mu} (\bar\psi\gamma_{\mu} \gamma_5 \psi +i\phi^{\ast}
D_{\mu}^{\leftrightarrow}\phi) =0\ ,}
and
\eqn\yav{\del_{\mu}
J^{\mu}=\del_{\mu}\bar\psi\gamma^{\mu}\psi=
{1\over 2\pi} \epsilon_{\mu\nu}F^{\mu\nu}\ .}
A simple path integral manipulation
relates the current equations to
Ward idendities for $\JJ$ .  One finds that
\eqn\wardone{
{\del\over\del y^{\alpha}} \JJ =0
\ ,}
and
\eqn\wardtwo{
\eqalign{{\del\over\del x^{\mu}}& \JJ =\cr
{1\over \pi}&
\epsilon^{\mu\alpha}{\del\over\del x^{\mu}}
\delta (x-y)+{1\over \pi}\bra{0} T^{\ast}
\left(\epsilon_{\mu\nu}\del^{\mu} A^{\nu}(x)
\  J^{\alpha}_5 (y)\right) \ket{0}\ .\cr}}
\smallskip
\noindent If it were not for the last term in ~\wardtwo , the
two Ward identities ~\wardone\ and ~\wardtwo\ would ensure
the existence
of a massless pole in the current correlator~\ref\???{A.D. Dolgov
and V.I. Zakharov,
Nucl. Phys. B27 (1971) 525\semi
Y. Frishman,  A. Schwimmer, T. Banks and S. Yankielowicz,
Nucl. Phys. B177 (1981) 157\semi
S. Coleman and B. Grossman, Nucl. Phys. B203 (1982) 205.}.
Naively one might expect the last term
in~\wardtwo\ to give at most ${\cal O}(g)$ perturbative corrections
to this pole or its residue.

We are thus confronted with the same dilemma as before.  The massive
spectrum of a spontaneously broken U(1) axial gauge theory appears to
be inconsistent with its vector current anomaly.  The
resolution of the puzzle lies in the fact that the
gauge boson mass is proportional to $g$.
It turns out that the last term in ~\wardtwo\ contains
an order zero piece which exactly cancels the first term at small
$p^2$.  The last term in ~\wardtwo\ can be rewritten as
\eqn\row{
-{1\over 2\pi}\bra{0}
T^*\left(\epsilon_{\mu\nu}\del^{\mu}A^{\nu}(x)\ 2v^2\left( \del^{\alpha}
\theta(y) + 2A^{\alpha}(y)\right)\right)\ket{0}\ ,}
where $\phi=\rho\exp{i\theta\gamma^5}$, $\bra{0}\rho\ket{0}=v$,
and terms which
do not give a zeroth order contribution have been dropped.
In  t'Hooft $\xi$-gauge there is no mixing between $\theta$ and
$A^{\mu}$,  so in momentum space the leading term of ~\row\ is
\eqn\pol{\eqalign{&{1\over \pi}
4v^2 \epsilon_{\mu\nu}
p^{\mu}g^2 ({g^{\nu\alpha}+ {(1-\xi)p^{\nu}p^{\alpha}\over
\xi p^2 - 4g^2 v^2}\over p^2 -4g^2 v^2})\cr
=& {1\over \pi}\epsilon^{\mu\alpha}p_{\mu} {4v^2g^2\over p^2- 4g^2
v^2}\ .\cr}}
At
small $p^2$ this is just $-{1\over \pi}\epsilon^{\mu\alpha}p_{\mu}$,
giving the stated cancellation.

An almost identical cancellation occurs in the Schwinger
model~\ref\schwing{J. Schwinger,  Phys. Rev. 128 (1962) 2425.}
with no fermion
mass term.
This model
also has a massive spectrum.  Furthermore the Ward identities are
like
those of the axial Higgs model, except that axial and vector labels are
swapped:
\eqn\taka{{\del\over \del y^{\alpha}}\JJQ =0\ ,}
and
\eqn\tosh{
\eqalign{{\del\over\del x^{\mu}}& \JJQ =\cr
-{1\over \pi}&
\epsilon^{\mu\alpha}{\del\over\del x^{\mu}}
\delta (x-y)-{1\over \pi}\bra{0} T^{\ast}
\left(\epsilon_{\mu\nu}\del^{\mu} A^{\nu}(x)
\ J^{\alpha} (y)\right) \ket{0}\ .\cr}}
\smallskip
\noindent In bosonized
form~\ref\mant{J. Frohlich and E. Seiler,  Helv. Phys. Acta.
49 (1976) 889\semi
N. S. Manton, Annals of Physics 159
(1985), 220.}
the last term of the latter ward identity can be written as
\eqn\bzz{\bra{0} T^{\ast} {e^2\over\pi\sqrt{\pi}}\phi(x)
\ {1\over\sqrt{\pi}}\epsilon^{\alpha\nu}\del_{\nu}\phi\ket{0}\ ,}
where $\phi$ is a scalar field with mass ${e\over\sqrt\pi}$.
At momentum small compared to the coupling $e$, this becomes
${1\over\pi}\epsilon^{\alpha\nu}p_{\nu}$ which cancels
against the first (anomalous commutator) term of~\tosh\ .
Thus the anomaly equation does not imply a massless pole.

\newsec{Conclusion}

The apparent paradox of an anomaly equation which is insensitive to
particle masses has been resolved in $1+1$ dimensions.  The Higgs mechanism
creates a gap, but also
provides a means to cross the gap.  In the presence of
a localized background with Pontryagin number one,  there is a bound
fermion
due to the winding Higgs background.  This bound fermion
acts as an ``elevator'' which carries charge across the gap.
For uniform charge generating backgrounds, the Higgs
degree of freedom prevents the
existence of an adiabatic limit.  In the dynamical case, the
gauge boson becomes massive due to the Higgs.  The
gauge boson mass alters
the anomalous ward identities in such a way that they do not
imply the existence of a massless state. I believe the mechanisms
described here should
extend readily to $3+1$ dimensions and the standard model.
\bigskip
\centerline{{\bf Acknowledgements}}
\bigskip
The author would like to thank David Kaplan, Aneesh Manohar and
Jan Smit for useful discussions. This work was supported in part by the
Department of Energy under grant number DOE-FG03-90ER40546, the Texas
National Research Laboratory Commission under grant RGFY93-206, and by
the National Science Foundation under grant PHY-8958081.

\newsec{Appendix A}
\leftline{{\bf Bogoliubov transformations for gauge theories: a paradox
}}
\leftline{{\bf with}
${\bf \epsilon^{\mu\nu}\gamma_{\nu}=\gamma^{\mu}\gamma^5}$}
\bigskip

In $1+1$ dimensions $\gamma^{\mu}\gamma^5= \epsilon_{\mu\nu}\gamma^{\nu}$
Therefore the $1+1$ dimensional Dirac equation with
an axial gauge field $A^{\mu}$ is equivalent to the
Dirac equation with a background vector gauge field $V^{\mu}$
where $V^{\mu}=\epsilon^{\mu\nu}A_{\nu}$.  Thus it naively appears that
if an axial gauge theory does not conserve vector charge, then
neither does
a vector gauge theory.  Conversly if a vector theory does not
conserve axial charge,  it seems that an axial theory does not
conserve axial charge either.  Fortunately both these statements
are not true.

The reason they are
not true in a finite volume is that there is an ambiguity
in doing Bogoliubov transformations.  This ambiguity is removed by
choosing either axial or vector gauge invariance.
Consider the massless
axial gauge theory in an $S_1\otimes R_1$ space-time,  and
suppose charge is
produced by a
field strengh which vanishes at asymptotic times.  The
change in vector charge is equal to minus the change in the
Chern-Simons number:
\eqn\chsim{\Delta Q={g\over\pi}\Delta\oint dx^1A_1\ .}
Therefore the gauge can be chosen so that $A^{\mu}$ vanishes
in either the asymptotic past or the asymptotic future, but not
both.  I will call the Fermi field $\psi^{in}$ or $\psi^{out}$
depending on whether $A^{\mu}$ vanishes in
the past or future.  $\psi^{in}$ can be expanded
in terms of spinors which have definite momentum and frequency
in the asymptotic past.
Similiarly $\psi^{out}$ can be expanded in terms of spinors which
have definite momentum and frequency in the asymptotic future.
Particle production is then determined from the Bogoliubov
transformation relating the two sets of expansion coefficients.

Now suppose we were to consider the vector gauge theory with the
backgrounds $V^{\mu}=\epsilon^{\mu\nu}A_{\nu}$.  Suppose also that
both the axial and vector field strenghths vanish at past and future
times.
If both field strengths vanish then
$\epsilon^{\mu\nu}\del_{\mu}A_{\nu}$ and
$\del_{\mu}A^{\mu}$ vanish and  $A^{\mu}$  must be a constant.
Consider a configuration with $A^{\mu}$=0 in that past and
$A^{\mu}=a^{\mu}$ in the future.  The difference between an
axial gauge theory and a vector gauge theory lies in the relation
between $\psi^{in}$ and $\psi^{out}$.  For the axial theory
\eqn\crung{\psi^{out}=\exp(iga_{\mu}x^{\mu}\gamma^5)\psi^{in}\ ,}
while for the vector theory
\eqn\smag{\psi^{out}=\exp(ig\epsilon_{\mu\nu}a^{\mu}x^{\nu})\psi^{in}\ .}
In light-cone coordinates,
the two $\psi^{out}$ fields are related by the transformation
\eqn\trog{\psi \rightarrow \exp(iga_+ x^+ P_L+iga_- x^- P_R)\psi\ .}
This transformation changes the vector charge by an amount
$g(a_+ - a_-){L\over 2\pi}$  and the axial charge by an amount
proportional to $g(a_+ + a_-){L\over 2\pi}$, where $L$ is
circumference of $S_1$.  Thus in a finite volume one finds the
desired result that the axial theory produces
only vector charge and the vector theory produces only axial charge.

The arguments above are not sufficient to show this result in an
infinite volume.  This is because in an infinite volume one can
always find a gauge in which the vector potential vanishes in both
the asymptotic past and asymptotic future \footnote*{In a finite volume
one is prevented from doing this by the gauge invariance of
$\exp(ig\oint dx^1A_1)$}.
For these gauges there is no difference between the out fields in
the axial theory and the out fields in the vector theory: both are
equal to the in field.
However there  is no equivalence between
localized gauge invariant backgrounds in the axial theory and
localized gauge invariant backgrounds in the vector theory provided
that either vector charge or axial charge respectively are produced.
If the axial and vector field strenghs are both localized, then
$\epsilon^{\mu\nu}\del_{\mu}A^{\nu}$ and $\del_{\mu}A^{\mu}$
vanish outside some finite region of space-time.  This means that
$A^{\mu}$ must be a constant outside this region.  The Pontryagin
index for both the axial and the vector theory therefore vanishes.
Note also that for the massive axial theory, a winding Higgs background
has no Q.E.D. counterpart.

\newsec{Appendix B}
\leftline{{\bf A No Pairing Theorem}}
\bigskip
The Euclidean equations of motion for the fermions of a spontaneously
broken axial gauge theory possess the symetry $\psi\rightarrow\
\gamma^0\psi^*$.  In this appendix I show that, in an adiabatic
limit, this symetry does not
yield independent solutions.  To be precise,  a solution of
$\hat K f_0(x,\tau)=0$ has the property that
$\gamma^0f_0^*(x,\tau)=\exp(i\alpha)f_0(x,\tau)$,
where the phase $\alpha$ is a constant.  The same is true for spinors
annihilated by the adjoint operator $\hat K ^{\dagger}$.
Recall that the solution of $\hat K f_0(x,\tau)=0$
in an adiabatic limit is
\eqn\adabatic{f_0(x,\tau)=
\exp\left(\int_0^{\tau}d\tau^{\prime}E_0(\tau^{\prime})\right)
\exp(i\beta(\tau))\chi_0(x,\tau)\ ,} where $\chi_0(x,\tau)$ is an
eigenfunction of
the time dependent Hamiltonian for which the energy $E_0(\tau)$ crosses
the gap.  The Berry's phase $\beta(\tau)$ will turn out to be important
to prevent pairing of zeromodes.   At asymptotic positive $x$ the
magnitude of the Higgs field is $v$, and one can always
choose the gauge so that the phase of the Higgs field is independent of
$x$.  With this choice the bound state eigenfunctions of $H(\tau)$ at
large $x$ are of the form \eqn\boust{\chi_0(x,\tau)=
\pmatrix{e^{ic(\tau)}\cr
         e^{-ic(\tau)}{E_0(\tau)+i\kappa(\tau)\over \lambda v}\cr}
e^{-\kappa(\tau)x}e^{ia(\tau)}\ ,}
where
\eqn\engci{\kappa(\tau)=\sqrt{\lambda^2v^2-E_0^2(\tau)}\ ,}
$c(\tau)$ is the phase of Higgs, and $a(\tau)$ is
an arbitrary phase.
Therefore at large
positive $x$
\eqn\lex{\gamma^0 \chi_0^*(x,\tau)= e^{-2ia(\tau)}
\sqrt{{E_0(\tau)-i\kappa(\tau)\over E_0(\tau)+i\kappa(\tau)}}
\ \chi_0(x,\tau)\ .}
It is easy show that the above
relation holds at all $x$ without knowing the exact form of the
solution.  If $\chi$ is an solution of \eqn\egval{(\hat H - E)\chi=0\ ,}
then so is $\gamma^0\chi^{\ast}$,  because \eqn\fnerg{\gamma^0\hat H ^{\ast}
\gamma^0= \hat H\ .}  Furthermore the eigenvalue
equation~\egval\  is linear and first order in $x$.  Therefore
if the relation~\lex\  is true at any $x$, then it must be true at
all $x$.  We thus arrive at the result that
\eqn\fullres{\gamma^0f_0^*(x,\tau)=
e^{-2i\beta(\tau)}e^{-2ia(\tau)}\sqrt{{E(\tau)-i\kappa(\tau)\over
E(\tau)+i\kappa(\tau)}}\, f_0(x,\tau)}
It appears that there is a time dependent phase relation, but in fact
the product of all the phases above is independent of $\tau$.
The Euclidean equations of motion are linear and first order in $\tau$,
and possess the symetry $\psi\rightarrow\gamma^0\psi^{\ast}$.
Therefore if at some fixed $\tau$
\eqn\ngan{\gamma^0f_0^*(x,\tau)=
e^{i\alpha}f_0(x,\tau)\ ,}
then this relation must hold at all $\tau$.
The symetry which gives pairs of
zeromodes
in the massless theory fails to give pairs in the spontaneously broken
theory.

\listrefs
\listfigs
\insertfig{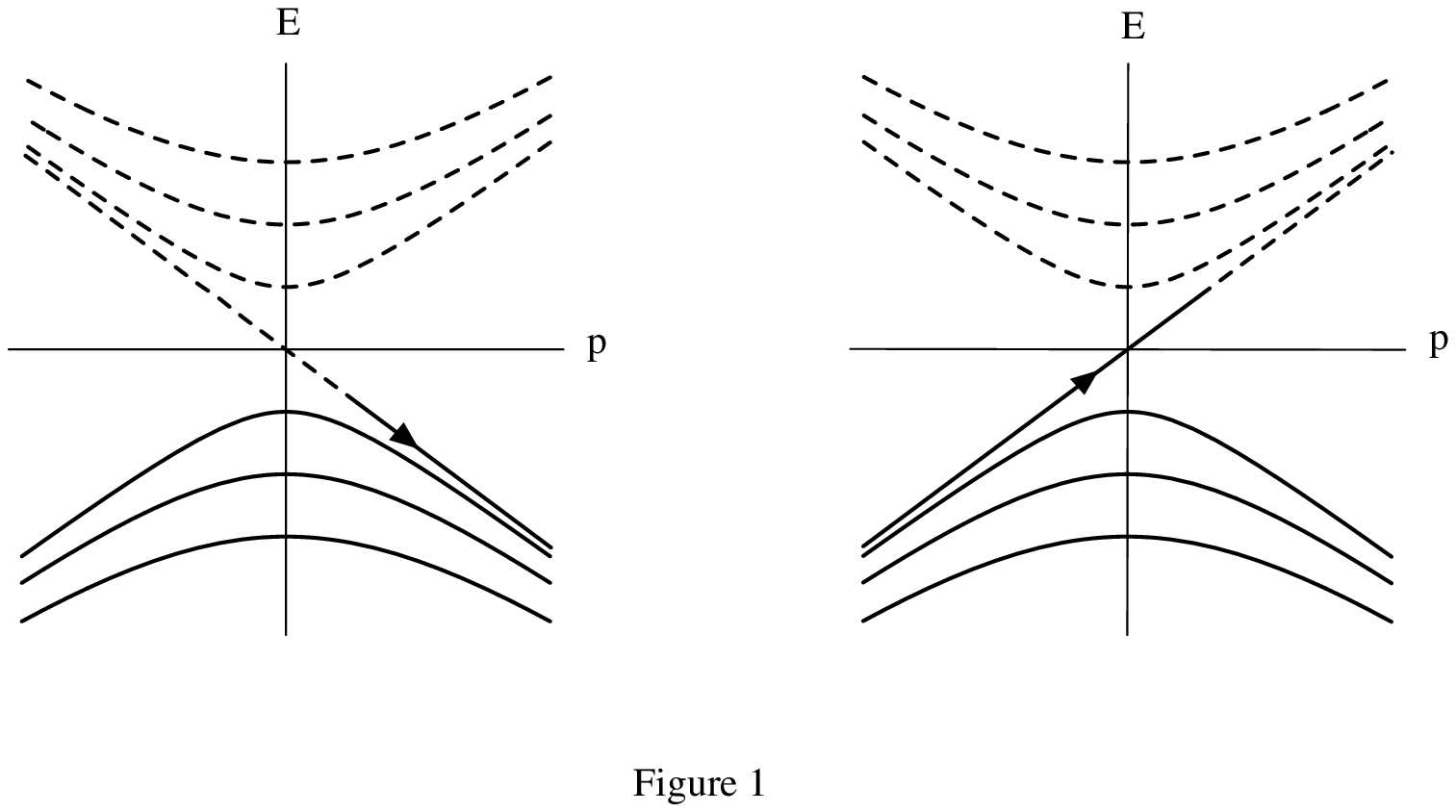}
\insertfig{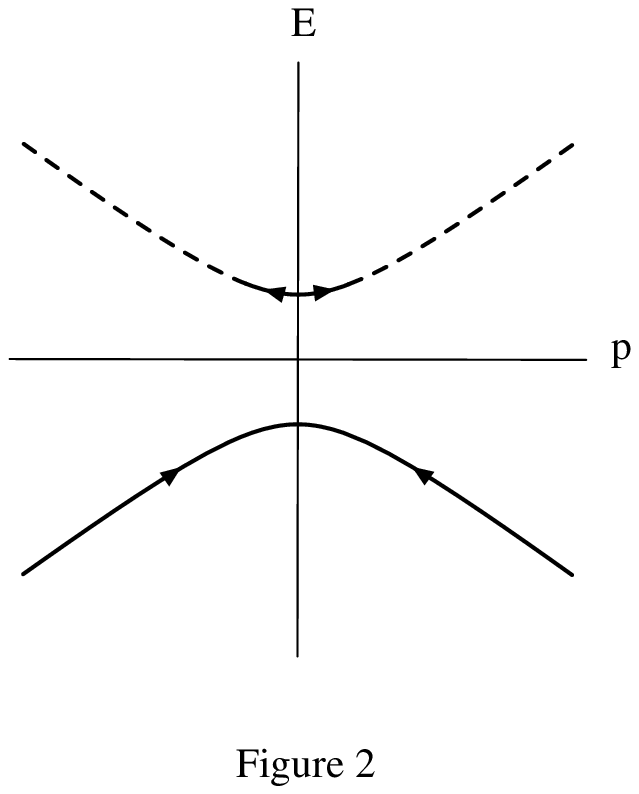}
\insertfig{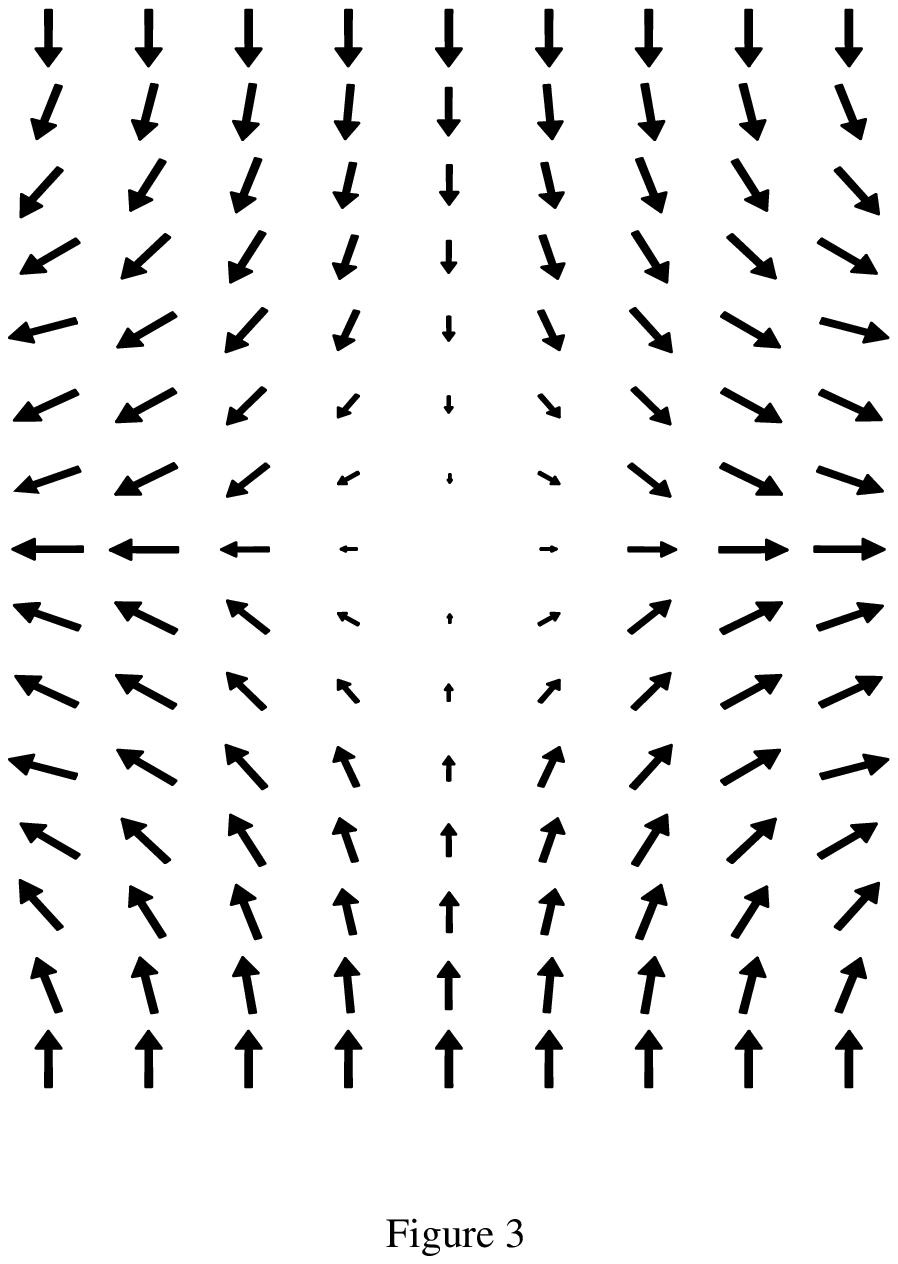}

\bye